\def \VersionWithComments {}

\documentclass{llncs}

\usepackage{amsmath}
\usepackage{amssymb}
\usepackage{stmaryrd}
\usepackage{graphicx}
\usepackage[ruled,vlined]{algorithm2e}
\usepackage{threeparttable}
\usepackage{extarrows}
\usepackage{mathrsfs}
\usepackage{stackrel}
\usepackage{proof}
\usepackage{booktabs}

\usepackage{mathtools}
\usepackage{parcolumns}
\usepackage{scalerel}
\usepackage{color}
\usepackage{bussproofs}
\usepackage{stackengine}
\usepackage{graphicx}
\usepackage{float}
\usepackage{tikz}

\usepackage{hyperref}
\DeclarePairedDelimiter{\ceil}{\lceil}{\rceil}
\usepackage{url}

\usepackage{pgf}
\usepackage{tikz}
\usetikzlibrary{arrows,automata,shapes}
\usepackage[latin1]{inputenc}
\usepackage{amssymb}
\def\kname{{$\mathbb{K}$}}

\graphicspath{{Figures/}}

\ifdefined \VersionWithComments
\usepackage{marginnote}
\newcommand{\marginX}{\marginnote{\huge{\quad\quad\textbf{!}\quad\quad}}}
\newcommand{\jj}[1]{\mbox{}{\color{green!50!black}\marginX{}\textbf{[Jiao}: #1]}}
\newcommand{\ksl}[1]{\mbox{}{\color{green!50!blue}\marginX{}\textbf{[Shuanglong}: #1]}}
\newcommand{\lsw}[1]{\mbox{}{\color{orange}\marginX{}\textbf{[Shang-Wei}: #1]}}
\newcommand{\instructions}[1]{{\color{red}\marginX{}\textbf{[Instructions: ``#1'']}}}
\newcommand{\reviewer}[2]{\mbox{}{\color{red}\marginX{}\textbf{[Reviewer #1}: ``#2'']}}
\newcommand{\todo}[1]{\mbox{}{\color{blue}{\marginX{}\textbf{TODO}\ifx#1\\\else:\ \fi #1}}} 
\else
\newcommand{\instructions}[1]{}
\newcommand{\jj}[1]{}
\newcommand{\ksl}[1]{}
\newcommand{\lsw}[1]{}
\newcommand{\reviewer}[2]{}
\newcommand{\todo}[1]{}
\fi

\usepackage{listings}
\newcommand{\chuhao}{\fontsize{8pt}{\baselineskip}\selectfont}
\def\code#1{\texttt{\chuhao#1}}
\def\codec#1{\texttt{#1}}
\def\rulename#1{\textsc{\textcolor{blue}{#1}}}
\def\kname{{$\mathbb{K}$}}
\newcommand\RWrule[3]{
\scaleleftright[1ex]{<}{
\begin{tabular}{c}
#1\\
\cline{1-1}
#2\\
\end{tabular}\,$\ldots$}
{>}$_{#3}$
}

\newcommand\Newrule[2]{
\begin{minipage}{1.0\textwidth}
~\\
{\small RULE \rulename{#1}}\\
{\footnotesize #2}\\
\end{minipage}
}

\newcommand\Rrule[2]{
\scaleleftright[1ex]{<}{$\ldots$\,
\begin{tabular}{c}
#1\\
\end{tabular}\,$\ldots$}
{>}$_{#2}$
}

\newcommand\RruleM[2]{
\scaleleftright[1ex]{<}{
\begin{tabular}{c}
#1\\
\end{tabular}} 
{>}$_{#2}$
}

\newcommand\RruleS[2]{
\scaleleftright[1ex]{<}{
\begin{tabular}{c}
#1\\
\end{tabular}\,$\ldots$}
{>}$_{#2}$
}

\begin{document}

\title{Executable Operational Semantics of Solidity}

\author{Jiao Jiao\inst{1} \and Shuanglong Kan\inst{1} \and Shang-Wei Lin\inst{1} \and David San\'{a}n\inst{1} \and Yang Liu\inst{1} \and Jun Sun\inst{2}}

\institute{
School of Computer Science and Engineering, Nanyang Technological University 
\and
Singapore University of Technology and Design 
}

\maketitle

\begin{abstract}
Bitcoin has attracted everyone's attention and interest recently. Ethereum (ETH), a second generation cryptocurrency, extends Bitcoin's design by offering a Turing-complete programming language called Solidity to develop smart contracts. Smart contracts allow creditable execution of contracts on EVM (Ethereum Virtual Machine) without third parties. Developing correct smart contracts is challenging due to its decentralized computation nature. Buggy smart contracts may lead to huge financial loss. Furthermore, smart contracts are very hard, if not impossible, to patch once they are deployed. Thus, there is a recent surge of interest on analyzing/verifying smart contracts. While existing work focuses on EVM opcode, we argue that it is equally important to understand and define the semantics of Solidity since programmers program and reason about smart contracts at the level of source code. In this work, we develop the structural operational semantics for Solidity, which allows us to identify multiple design issues which underlines many problematic smart contracts. Furthermore, our semantics is executable in the K framework, which allows us to verify/falsify contracts automatically.
\end{abstract}

\section{Introduction} \label{sec:Introduction}

The success of Bitcoin since 2009 stimulates the development of other blockchain based applications such as Ethereum. Ethereum is a second generation of cryptocurrency which supports the revolutionary idea of smart contracts. A smart contract~\cite{Smartcon} is a computer program written in a Turing-complete programming language called Solidity, which is stored on the blockchain to achieve certain functionality. Smart contracts benefit from the features of blockchain in various aspects. For instance, it is not necessary to have an external trusted authority in order to achieve consensus, and transactions through smart contracts are always traceable and credible.

Smart contracts must be verified for multiple reasons. Firstly, due to the decentralized nature of blockchain, smart contracts are different from original programs (e.g., C/Java). For instance, in addition to stack and heap, smart contracts operate a third `memory' called storage, which are permanent addresses on the blockchain. Programming smart contracts thus is error-prone without a proper understanding of the underlying semantic model. This is further worsened by multiple language design choices (e.g., fallback functions) made by Solidity. In the following, we illustrate the difference between smart contracts and original programs using the contract named \texttt{Test2} shown in Fig.~\ref{fig:StrangeProgram}~{(b)}, where \texttt{b} is a global two dimension array, i.e., \texttt{b[0] = [1,2,3]} and \texttt{b[1] = [4,5,6]}. In function \texttt{foo2()}, a local array \texttt{d} of three elements is declared. Its second and third elements are set to be $10$ and $11$ respectively afterwards. After the execution of \texttt{foo2()},  the global array \texttt{b} is changed as: \texttt{b[0] = [0,10,0]} and \texttt{b[1] = [4,5,6]}. To understand this surprising behavior, we must understand the storage/memory model of Solidity, and make sure it is formally defined so that programmers can write contracts accordingly. If a programmer implements a smart contract with his/her intension inconsistent with the Solidity semantics, vulnerabilities are very likely to be introduced.

\begin{figure}[tb]
\scriptsize
\centering
\begin{minipage}{0.35\linewidth}
\centering
\begin{lstlisting}[frame=single,numbers=left,numberstyle=\tiny]
contract Test {
   uint128 a = 1;
   uint256 b = 2;

   function foo() public {
      uint256[2] d;
      d[0] = 7;
      d[1] = 8;
   }
}
\end{lstlisting}
\vspace{-1mm}(a)
\end{minipage}
~ \hspace{8mm}
\begin{minipage}{0.5\linewidth}
\centering
\begin{lstlisting}[frame=single,numbers=left,numberstyle=\tiny]
contract Test2 {
   uint128 a = 9;
   uint128[3][2] b = [[1,2,3],[4,5,6]];

   function foo2() public {
      uint256[3] d;
      d[1] = 10;
      d[2] = 11;
   }
}
\end{lstlisting}
\vspace{-1mm}(b)
\end{minipage}
\caption{Strange Contracts} \label{fig:StrangeProgram}
\end{figure}
%

Secondly, a smart contract can be created and called by any user in the network. A bug in the contract potentially leads to threats to the security properties of smart contracts. Verifying smart contracts against such bugs is crucial for protecting digital assets. One well-known attack on smart contracts is the DAO attack~\cite{DAO}. The attacker exploited a vulnerability associated with fallback functions and the reentrancy property in the DAO smart contract~\cite{DBLP:conf/post/AtzeiBC17}, and managed to drain more than 3.6 million ETH (i.e., the Etheruem coin which has a value of about \$1000 at the time of writing). Thirdly, unlike traditional software which can be patched, it is very hard if not impossible to patch a smart contract, once it is deployed on the blockchain due to the very nature of blockchain. For instance, the team behind Ethereum decided to conduct a soft-fork of the Ethereum network in view of the DAO attack, which caused a lot of controversial. It is thus extremely important that a smart contract is verified before it is deployed as otherwise it will be forever under the risk of being attacked.


There have been a surge of interests in developing analysis/verification techniques for smart contracts~\cite{DBLP:conf/ccs/LuuCOSH16,KEVM,Amani:2018:TVE:3176245.3167084,10.1007/978-3-319-70278-0_33}. For instance, the authors in~\cite{DBLP:conf/ccs/LuuCOSH16} developed a symbolic execution engine called Oyente which targets bytecode running on Ethereum Virtual Machine (EVM). Since Solidity programs are compiled into bytecode and run on EVM, Oyente can be used to analyze Solidity programs. In addition, the authors in~\cite{KEVM} developed a semantic encoding of EVM bytecode in the K-framework. To the best of our knowledge, all existing approaches focus on bytecode. We believe that it is equally important to formally understand the semantics of Solidity since programmers program and reason about smart contracts at the level of source code. Otherwise, programmers are required to understand how Solidity programs are compiled into bytecode in order to understand them, which is far from trivial. 

In this work, we develop the structural operational semantics (SOS) for the Solidity programming language so that smart contracts written in Solidity can be formally reasoned about. The contributions of this work are twofold. Firstly, our work is the first approach, to our knowledge, on the formal semantics of the Solidity programming language other than the Solidity compiler itself. Our executable semantics covers most of the semantics specified by the official Solidity documentation~\cite{Sol}. Secondly, we implement the proposed SOS in K-framework~\cite{K}, which provides a Reachability Logic prover~\cite{oopsla/StefanescuPYLR16}. With the proposed SOS and its implementation, we are able to detect vulnerabilities or reason about the correctness of smart contracts written in Solidity systematically.

The remaining part of this paper is organized as follows. In Section~\ref{sec:Background}, we introduce the background of Solidity smart contracts. The proposed executable operational semantics is introduced in Section~\ref{sec:SoliditySemantics}. In Section~\ref{sec:SemanticsInK}, we introduce the implementation of Solidity semantics in K-framework by illustrating some important rules. Section~\ref{sec:Evaluation} shows the evaluation results of our Solidity semantics in K-framework. In Section~\ref{sec:RelatedWorks}, we review related works. Section~\ref{sec:Conclusion} concludes this work and discusses our future directions.


\section{Background of Solidity Smart Contracts} \label{sec:Background}

Ethereum, proposed in late $2013$ by Vitalik Buterin, is a blockchain-based distributed computing platform supporting smart contract functionality. It provides a decentralized international network where each participant node (also known as \textit{miner}) equipped with EVM can execute smart contracts. Ethereum also provides a cryptocurrency called ``ether'' (ETH), which can be transferred between different accounts and used to compensate participant nodes for their computations on smart contracts.

Solidity is one of the high-level programming languages to implement smart contracts on Ethereum. A smart contract written in Solidity can be compiled into EVM bytecode and then be executed by any participant node equipped with EVM. Fig.~\ref{fig:SmartContractExample} shows an example of Solidity smart contract, named \texttt{Coin}, implementing a very simple cryptocurrency. A Solidity smart contract is a collection of code (its functions) and data (its state) that resides at a specific address on the Ethereum blockchain. In line~$2$, the public state variable \texttt{minter} of type \texttt{address} is declared to store the minter of the cryptocurrency, i.e., the owner of the smart contract. The constructor \texttt{Coin()}, which has the same name as the smart contract, is defined in lines~$5$--$7$. Once the smart contract is created and deployed\footnote{How to create and deploy a smart contract is out of scope and can be found in: \url{https://solidity.readthedocs.io/}}, its constructor is invoked automatically, and \texttt{minter} is set to be the address of its creator (owner), represented by the built-in keyword \texttt{msg.sender}. In line~$3$, the public state variable \texttt{balances} is declared to store the balances of users. It is of type \texttt{mapping}, which can be considered as a hash-table mapping from keys to values. In this example, \texttt{balances} maps from an user (represented as an address) to his/her balance (represented as an unsigned integer value).

The \texttt{mint()} function, defined in lines~$9$--$12$, is supposed to be invoked only by its owner to mint \texttt{amount} coins for the user located at the \texttt{receiver} address. If \texttt{mint()} is called by anyone except the owner of the contract, nothing will happen because of the guarding \texttt{if} statement in line~$10$. The \texttt{send()} function, defined in lines~$14$--$18$, can be invoked by any user to transfer \texttt{amount} coins to another user located at the \texttt{receiver} address. If the balance is not sufficient, noting will happen because of the guarding \texttt{if} statement in line~$15$; otherwise, the balances of both sides will be updated accordingly.
\begin{figure}[tb]
\scriptsize
\centering
\begin{minipage}{0.8\linewidth}
\centering
\begin{lstlisting}[frame=single,numbers=left,numberstyle=\tiny]
contract Coin {
   address public minter;
   mapping (address => uint) public balances;

   function Coin() public {
      minter = msg.sender;
   }

   function mint(address receiver, uint amount) public {
      if (msg.sender != minter) return;
      balances[receiver] += amount;
   }

   function send(address receiver, uint amount) public {
      if (balances[msg.sender] < amount) return;
      balances[msg.sender] -= amount;
      balances[receiver] += amount;
   }
}
\end{lstlisting}
\end{minipage}
\caption{Smart Contract Example} \label{fig:SmartContractExample}
\end{figure}

A blockchain is actually a globally-shared transactional database or ledger. Every participant node can read the information in the blockchain. If one wants to make any state change in the blockchain, he or she has to create a so-called \textit{transaction} which has to be accepted and validated by all other participant nodes. Furthermore, once a transaction is applied to the blockchain, no other transactions can alter it. For example, deploying the \texttt{Coin} smart contract generates a transaction because the state of the blockchain is going to be changed, i.e., one more smart contract instance will be included. Similarly, any invocation of the functions \texttt{mint()} and \texttt{send()} also generates transactions because the state of the contract instance, which is a part of the whole blockchain, is going to be changed. As mentioned earlier, each transaction has to be validated by other participant nodes. This validation procedure is so-called \textit{mining}, and the participant nodes validating transactions are called \textit{miners}. To motivate the miners to execute and validate transactions on the blockchain, they are rewarded ETH, which drives the blockchain functionally. Thus, each transaction is charged with a transaction fee to reward the miners. The transaction fee depends on the product of the gas price and the amount of gas. When one requests a transaction to be mined, he or she can decide the gas price to be paid. Of course, the higher the price is, the more likely miners are willing to validate the transaction. The amount of gas is determined by the computation required by the transaction. The more computation a transaction requires, the more gas miners will charge.

A smart contract instance has two memory areas to store data, namely, {\em storage} and {\em memory}. Storage is an array of storage slots, each of which is $32$~bytes long and addressable by an $256$-bit address. When a contract is created, its static data is allocated in the storage from storage slot~$0$ and growing to higher slots. Dynamic data such as pushing a new element into a dynamic array is allocated based on a hash algorithm. Complex datatypes such as arrays or structs are aligned to storage slot ($32$~bytes), adding padding to fill the necessary space. Primitive datatypes allocated in adjacent positions are packed together to save space. The storage/memory model is formally introduced in Section~\ref{sec:SoliditySemantics}. 

\section{Formal Specification of Solidity} \label{sec:SoliditySemantics}
We build an abstract model of the Solidity semantics. These representations of rules omit the details in the rule formation in K-framework, but reveal the idea of the semantics from a general as well as abstract perspective.
\subsection{Notations} \label{subsec:Notation}



\subsubsection{Type Specification}

Types are inductively defined in $\mathbb{T}$, as follows:
\vspace{-1mm}
\[
\begin{array}{l}
  \mathbb{K} := \mathsf{uint_m}\ |\ \mathsf{address}\ |\ \mathbb{K}[n] \\
  \mathbb{T} := \mathsf{uint_m}\ |\ \mathsf{address}\ |\ \mathbb{T}[n] \,\, |\ \mathbb{T}[\,]\ |\ \mathsf{mapping}\  \mathbb{K}\ \mathbb{T} \,\,|\ \langle \mathbb{T} \cdots \mathbb{T} \rangle\ |\ \mathsf{contract} \mathbb{c} \,\, |\ \mathsf{ref}\ \mathbb{T}
\end{array}
\]

Implicitly, types in Solidity have associated a memory space indicating whether a variable, and hence an expression, is allocated in the storage or memory. To consider this, we extend the definition of type as a tuple $\mathbb{T} = (\Psi |\ M) \times \mathbb{T}$, where $\Psi$ refers to the storage and $M$ to the memory.

$\mathbb{T}$ includes usual data type constructors of high level languages such as arrays and dynamic arrays represented by $\mathbb{T}[n]$ and $\mathbb{T}[\,]$, respectively. Structures are represented by $\langle \mathbb{T} \cdots \mathbb{T} \rangle $. Keys for mappings are constrained to types included in type $\mathbb{K}$. The type $\mathsf{uint_m}$ defines $2^m$-bit unsigned integers for $3 \leq m \leq 8$. We do not provide specification for other primitive types other than $\mathsf{uint}$ because we can construct them based on $\mathsf{uint}$. The type $\mathsf{ref}$ is not an explicit type in Solidity, but the compiler uses it implicitly when allocating in memory variables with complex data types. Solidity includes specific types such as contracts allocated in the ledger represented by the type $\mathsf{contract}$.

\subsubsection{State Specification}
We represent the storage as $\Psi:\mathbb{P} \mapsto \mathbb{B}$, a function from addresses in the domain of positive numbers to bytes. Since static variables are allocated in the storage sequentially from address 0, we attach a next address $\Lambda$ to $\Psi$ to indicate where a new declared variable is allocated into. In addition, $\Psi$ is also associated with a name space $N_\Psi: \mathbb{ID}_\Psi \mapsto\mathbb{P}\cup None$ and a type space $\tau_\Psi: \mathbb{ID}_\Psi \mapsto\mathbb{T}\cup None$, respectively mapping variables to memory addresses and types. We denote with $\mathsf{id} \notin N_\Psi$ or $\mathsf{id} \notin \tau_\Psi$ to represent that $\mathsf{id}$ is mapped to $None$ in the name space or type space, respectively.

Memory, denoted by $M$, is also a function $M:\mathbb{P} \mapsto \mathbb{B}$ from addresses to bytes. It is used to store local variables in functions. Similar to storage, $M$ is also associated with a name space $N_M: \mathbb{ID}_M \mapsto \mathbb{P}\cup None$ and a type space $\tau_M:\mathbb{ID}_M \mapsto\mathbb{T}\cup None $. Differently from the spaces in the storage, the memory contains a stack of name and type spaces $\amalg$ to model new scopes when calling a function. For simplicity, $N_M$ and $\tau_M$ access the top spaces in $\amalg$ and we omit $\Psi$ and $M$ when it is clear in the context.

We use $\sigma = (\Psi, M, \Omega)$ to denote the storage and memory configuration of a smart contract instance, and $\Omega$ is a stack of storage configuration for handling external function calls. We access the elements of $\sigma$ with $\Psi_\sigma$, $M_\sigma$, and $\Omega_\sigma$. The overall configuration of a smart contract instance is denoted by $\langle \sigma, \mathsf{Prog} \rangle$, where $\mathsf{Prog}$ is the set of program statements. After a smart contract instance is deployed on the blockchain, it is identified by a unique $256$-bit address. We denote the configuration of the blockchain by $\Delta$, which is a function mapping $\alpha$, a $256$-bit address of a contract instance, to its configuration $\sigma$, i.e., $\Delta(\alpha) = \sigma$. 

To access locations in $\Psi$ and $M$, we use $[addr]_{store}^{size}$ to represent a value stored in position $addr$ of $size$ bytes stored in $store$ . Thus, given a variable $\mathsf{id}$ allocated in $\Psi$, the notation $[N_\Psi(\mathsf{id})]_{\Psi}^{Size\,(\tau_\Psi(\mathsf{id}))}$ represents the stored value of $id$. Function $Size$ gives the number of bytes a datatype, considering packing and padding applied in Solidity during variable allocation. We provide a detailed specification in Appendix~\ref{subsec:MVeval}. 

\subsubsection{State Modifications in Rules}

For a configuration $\sigma$, we use $\sigma'$ to denote the configuration after we apply changes on $\sigma$. For example, the following rule $\mathsf{Ex}$ does two changes in $\sigma$: (1) it modifies the type space $\tau$ in the storage $\Psi$  for the element $\mathsf{id}$ to value $\mathsf{Type}$, and (2) it modifies the name space $N$ in the storage $\Psi$  for the element $\mathsf{id}$ to value $\mathsf v$. Any other state component such as the storage $\Psi$, the memory $M$, or elements different from $\mathsf{id}$ in $\tau_\Psi$ and $N_\Psi$ are not changed.


\[
\footnotesize
\infer[\mbox{Ex}]{\langle \sigma, \mathsf{Exp} \rangle \longrightarrow \langle \sigma', \cdot \rangle }{%
	\begin{array}{l}	  
	\tau_{\Psi_{\sigma'}} \mathsf{(id) = Type} 
	\end{array}
    & \begin{array}{l}	  
    N_{\Psi_{\sigma'}} \mathsf{(id) = v} 
    \end{array}}
\]

%
%

\subsection{Structural Operational Semantics of Solidity} \label{subsec:SoliditySOS}

We abstract the semantics of solidity in rules for statements, expressions, and types. Most of the statements and semantics in Solidity are very similar to those used in high level languages like Java and C. For space reasons, we focus on the semantic for the evaluation of expressions involving storage/memory access, e.g., variable access, arrays, and mapping. We also focus on statements for variable declaration and function calls. We define the rules inductively based on Solidity type constructors and statements. 

We first start with the semantics for instructions defining variables represented by rules VD$_1$ and  VD$_2$. Notice that if a state variable is declared without giving the initial value, it is initialized as zero.

\[
\footnotesize
\begin{array}{lcr}
\infer[\mbox{VD}_1]{\langle \sigma, \mathsf{Type\,\,\, id = Exp \,;} \rangle \longrightarrow \langle \widehat{\sigma}', \cdot \rangle }{%
	\begin{array}{l}		
	E_R(\sigma, \mathsf{Exp}) \longrightarrow \langle \widehat{\sigma}, v \rangle \\
	\mathsf{N_{\widehat{\sigma}'} id = \ceil*{\Lambda_{\widehat{\sigma}}}^{(l, Type)}} \\ 
	\mathsf{{\Lambda}_{\widehat{\sigma}'} = \Lambda_{\widehat{\sigma}}\uparrow(l, Type)}
	\end{array}
	& \begin{array}{l}	  
	\mathsf{\tau_{\widehat{\sigma}'} id = Type} \\	
	\mathsf{id} \not\in N_{\widehat{\sigma}} \\	
	\mathsf{[N_{\widehat{\sigma}'} id]_{\widehat{\sigma}'}^{Size\ Type} =} v
	\end{array}}
&\hspace{5mm} &
\infer[\mbox{VD}_2]{\langle \sigma, \mathsf{Type\,\,\, id = Exp \,;} \rangle \longrightarrow \langle \widehat{\sigma}', \cdot \rangle }{%
	\begin{array}{l}		
	E_R(\sigma, \mathsf{Exp}) \longrightarrow \langle \widehat{\sigma}, v \rangle \\	 
    \mathsf{M_{\widehat{\sigma}'} = fr_{\widehat{\sigma}}(id,Type,}v)
	\end{array}
	& \begin{array}{l}
	\mathsf{id} \not\in N_{\widehat{\sigma}}
	\end{array}
}
\end{array}
\]

\begin{figure}[tb]
\centering
\begin{minipage}{1\linewidth}
\centering
\includegraphics[width=0.85\linewidth]{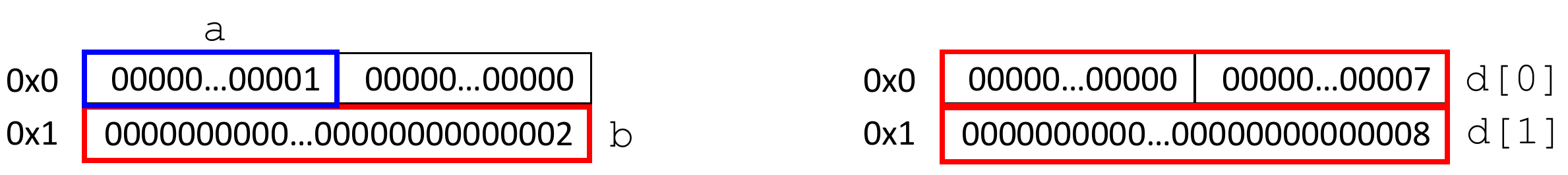}
\end{minipage}
\\
(a) \hspace{60mm} (b)
\\[1mm]
\begin{minipage}{1\linewidth}
\centering
\includegraphics[width=0.85\linewidth]{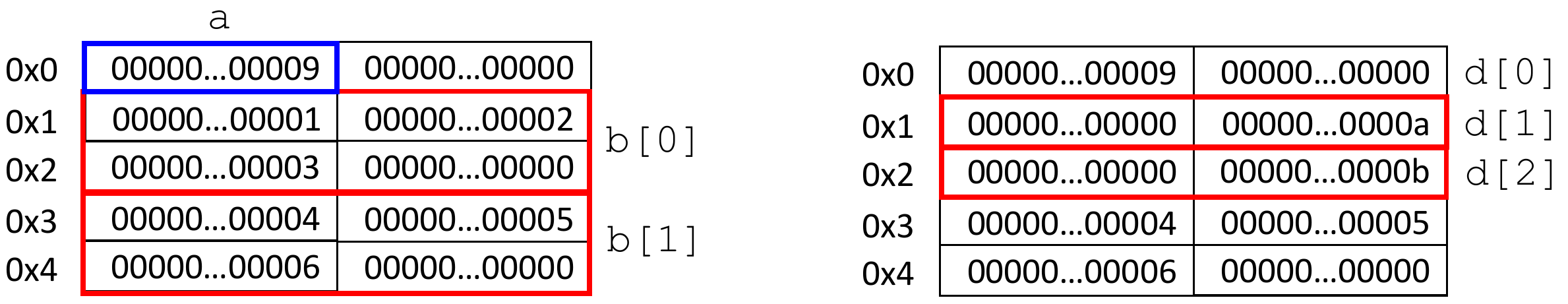}
\end{minipage}
\\
(c) \hspace{60mm} (d)
\caption{State Variable Declaration}
\label{fig:AllocExample}
\end{figure}

Let us take the example in Fig.~\ref{fig:StrangeProgram}~{(a)} to illustrate the rules for variable declarations. Here, we have two state variables declaration statements for \texttt{a} and \texttt{b}, respectively. To allocate \texttt{a}, the rule performs six steps. (1) the rule calculates the value \texttt{a} is going to be initialized to using the R-Value of $\mathsf{Exp}$ from state $\sigma$. Note that an expression can be a function, therefore the $\mathsf{Exp}$ may have side effects and modify $\sigma$ to $\widehat{\sigma}$. (2-3) the rule updates the name space and type space for variable $\mathsf{id}$. The type space is updated to $\mathsf{Type}$, i.e., \texttt{uint128} in this example. We update the name space using the function $\ceil*{\Lambda}^{l,Type}$. $\ceil*{\Lambda}^{l,Type}$ uses Solidity rules for allocation of variables to calculate if the declared variable must be allocated in the current free position $\Lambda$ or in the next position aligned to $l$. If the declared variable is a complex datatype then it must be allocated in a memory address aligned to $l$. Otherwise, it calculates the next address aligned to $\Lambda$ if the current position plus the size of the datatype is bigger than the next address aligned to $l$. In Solidity the memory alignment $l$ is $32$ bytes. In the example, $\Lambda$ takes the initial value of zero which is aligned to $32$. Note that $\mathsf{VD_1}$ refers only to storage variables, therefore we omit $\Psi$ in the rule when accessing $\mathsf{N}$ and $\tau$. (4), $\Lambda$ is updated using $\Lambda\uparrow(l, Type)$ first aligning $\Lambda$ to $l$ if it is necessary, and then it increases it with the size of the type of the variable, which in the example is $16$, the size of ($uint128$). The rules to calculate the size of a type considers padding and packing of types (see~\ref{subsec:MVeval} for more details). (5) the memory address where the variable is allocated takes the value of the R-Value of the expression, which is $1$. And (6) the rule checks that the current id does not belongs to the name space in $Psi$.  After the allocation, variable \texttt{a} is allocated in the beginning of slot~$0$ and occupies $16$ bytes, as shown in the blue box of Fig.~~\ref{fig:AllocExample}~{(a)}. 

For state variable \texttt{b}, its type is \texttt{uint256}, which is the basic type of $256$-bit unsigned integers. Since variable \texttt{b} requires $32$ bytes, we are not able to allocate \texttt{b} within the next address aligned to $32$, that is within the slot~$0$. Instead, we need to allocate it in the next storage slot. After the allocation, variable \texttt{b} is allocated at address $32$ (the beginning of slot~$1$) and occupies $32$~bytes, as shown in red box of Fig.~\ref{fig:AllocExample}~{(a)}. The auxiliary variable $\mathtt{\Lambda}$ is updated to $64$ accordingly.

Let see another smart contract example in Fig.~\ref{fig:StrangeProgram}~{(b)}. After allocating the state variable \texttt{a}, auxiliary variable $\Lambda$ become $16$, as shown in the blue part of Fig.~\ref{fig:AllocExample}~{(c)}. Notice that \texttt{b} is an array of two elements, each of which is an array of three \texttt{uint128} integers. That is, \texttt{b[0] = [1,2,3]} and \texttt{b[1] = [4,5,6]}. To allocate the state variable \texttt{b}, rule Size used in function $\Lambda\uparrow(32, \texttt{uint128[3][2]})$ packs together the first two unsigned integers of the first dimension of the array, and adds padding for the second one to align the array to $32$. Then the size of the first dimension of \texttt{b} is $64$ bytes and the total size of the type is $128$ bytes. So, totally, four storage slots are allocated to \texttt{b}, as shown in the red part of Fig.~\ref{fig:AllocExample}~{(c)}. 

The rule VD$_2$ is similar the rule for state variable declarations except that the target is memory instead of storage and that we allocate the position and the R-Value of the expression in a fresh location. Function \texttt{fr} updates \texttt{id} in spaces \texttt{N.M} and $\tau.\texttt{M}$ with a new fresh address in memory and \texttt{Type}, respectively. Additionally \texttt{fr} copies the expression R-Value to the new location. After all the state variables are declared, whenever they need to be evaluated in statements, the following two rules are applicable to get their L-values and R-values.

Rule E\_{RV} returns the value of address \texttt{addr} in the configuration $\widehat{\sigma}$ obtained after the l-valuation of expression \texttt{exp} in the state $\sigma$. The accessed memory space, i.e., \texttt{M} or $\Psi$, is calculated from the type of \texttt{exp}. This behaviour is abstracted in the function \texttt{ST}.

\[
\footnotesize
\infer[\mbox{E}_{RV}]{E_{R}(\sigma, \mathsf{exp}) \longrightarrow \langle \widehat{\sigma},v\rangle }{%
	\mathsf{E(\sigma, exp) \longrightarrow (\widehat{\sigma}, addr)}
	& \mathsf{[addr]_{(ST\ exp)_{\widehat{\sigma}}}^{Size (Type_{\widehat{\sigma}}\ exp)} =} v 
}
\]

\[
\footnotesize
\begin{array}{lcr}
\infer[\mbox{E-ID1}]{E(\sigma, \mathsf{id}) \longrightarrow \langle \sigma,  (N_{\Psi_{\sigma}} \mathsf{id}) ) \rangle}{%
	\mathsf{id} \in N_{\Psi_{\sigma}} \quad \mathsf{id} \notin N_{M_{\sigma}}
}
& \hspace{10mm} &
\infer[\mbox{E-ID2}]{E(\sigma, \mathsf{id}) \longrightarrow \langle \sigma,  (N_{\sigma_{M}} \mathsf{id}) ) \rangle}{%
	\mathsf{id} \in N_{M_{\sigma}}
}
\end{array}
\]

Let us see function \texttt{foo()} in Fig.~\ref{fig:StrangeProgram}~{(a)}. Here, we declare a local array \texttt{d}, whose type is \texttt{uint256[2] storage}. Notice that, in Solidity, if an array is declared in a function without specifying the area (either in storage or memory), its default area is in storage. However, since we did not initialized the array \texttt{d}, it references the storage slot $0$ by default. Thus, changing the values of \texttt{d[0]} and \texttt{d[1]} overwrites the content in storage slots~$0$ and $1$, as shown in the red part of Fig.~\ref{fig:AllocExample}~{(b)}. That is why the values of \texttt{a} and \texttt{b} are changed to $0$ and $8$ after the execution of function \texttt{foo()}. Similarly, the local array \texttt{d} in function \texttt{foo2()} in Fig.~\ref{fig:StrangeProgram}~{(b)} references storage slot~$0$ by default as well. Thus, changing \texttt{d[1]} and \texttt{d[2]} overwrites the content in storage slots~$1$ and $2$, as shown in the red part of Fig.~\ref{fig:AllocExample}~{(d)}. That is why the global array \texttt{b} becomes \texttt{[0,10,0]} after function \texttt{foo2()} is executed. The evaluation of arrays can be performed by the following rules inductively.

\[
\footnotesize
\infer[\mbox{E-ARRAY}]{E(\sigma, \mathsf{Exp_{b}[Exp_{i}]}) \longrightarrow \langle \widehat{\sigma}', v \rangle}{
	\begin{array}{l}
	\mathsf{Type_\sigma\ Exp_b = T[n]} \\
	\mathsf{i\leq n}
	\end{array}
	& \begin{array}{l}
	E_{R}(\sigma, \mathsf{Exp_{i}}) \longrightarrow \langle \widehat{\sigma}, i \rangle \\
	E(\widehat{\sigma}, \mathsf{Exp_b}) \longrightarrow \langle \widehat{\sigma}', addr_b \rangle \\
	v = \mathsf{addr_b + (i*Size\ T)}
	\end{array}
}
\]

\[
\footnotesize
\infer[\mbox{E-ARRAY-LEN}]{E(\sigma, \mathsf{exp.length}) \longrightarrow \langle \widehat{\sigma}, \mathsf{len}\rangle}{%
	\mathsf{Type_\sigma exp =  Type [\,]}
	& E_R(\sigma, \mathsf{Exp_b}) \longrightarrow \langle \widehat{\sigma}, \mathsf{len} \rangle
}
\]

\begin{figure}[tb]
\scriptsize
\centering
\begin{minipage}{0.4\linewidth}
\centering
\begin{lstlisting}[frame=single,numbers=left,numberstyle=\tiny]
contract Test3 {
   uint256[] a;
		
   function foo3() public {
      a.push(10);
	  a.push(11);
   }
}
\end{lstlisting}
\vspace{-1mm}(a)
\end{minipage}
~ \hspace{8mm}
\begin{minipage}{0.4\linewidth}
\centering
\begin{lstlisting}[frame=single,numbers=left,numberstyle=\tiny]
contract Test4 {
   mapping(uint=>uint) a;
		
   function foo4() public {
      a[100] = 10;
	  a[200] = 11;
   }
}
\end{lstlisting}
\vspace{-1mm}(b)
\end{minipage}
\caption{Dynamic Arrays and Mapping} \label{fig:ArrayMapExample}
\end{figure}

\begin{figure}[tb]
	\centering
	\begin{minipage}{1\linewidth}
		\centering
		\includegraphics[width=0.95\linewidth]{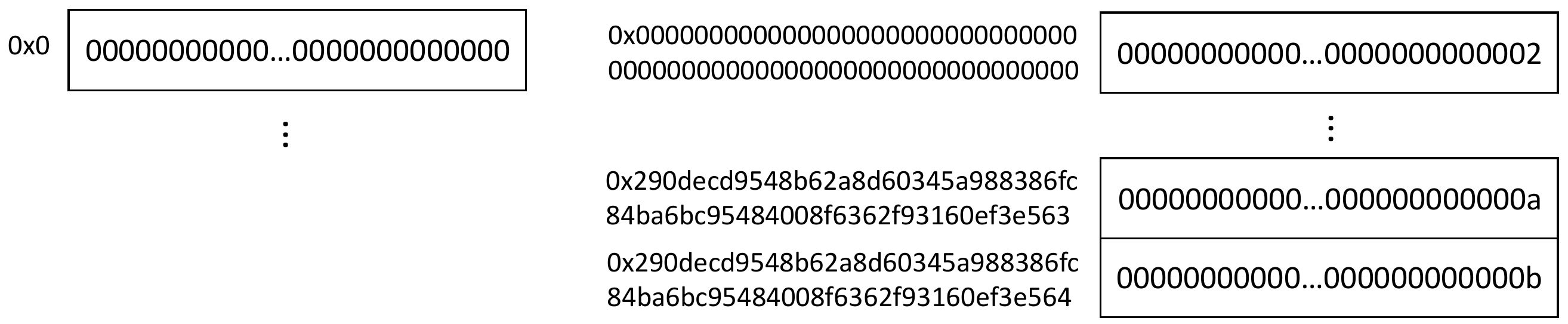}
	\end{minipage}
	\\[1mm]
	(a) \hspace{60mm} (b)
	\caption{Dynamic Array}
	\label{fig:DynamicArray}
\end{figure}

\begin{figure}[tb]
	\centering
	\begin{minipage}{1\linewidth}
		\centering
		\includegraphics[width=1\linewidth]{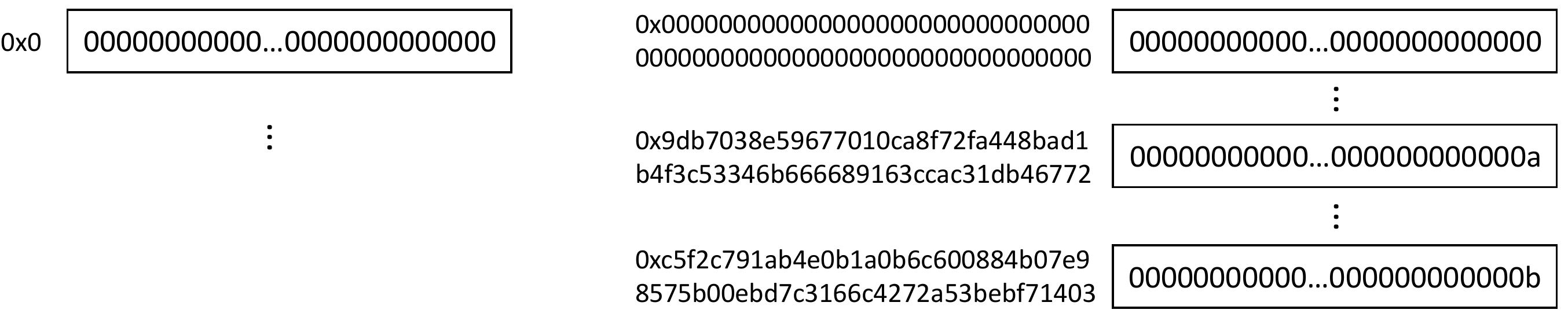}
	\end{minipage}
	\\[1mm]
	(a) \hspace{60mm} (b)
	\caption{Mapping}
	\label{fig:Mapping}
\end{figure}

In addition, Solidity provides two special data structures, dynamic array and mapping, which allocation is based on hash functions. Fig.~\ref{fig:ArrayMapExample}~{(a)} shows a simple contract using a dynamic array \texttt{a}. When a dynamic array is declared, it has no element, and one storage slot is allocated to it as the base slot to store the number of elements it has so far, as shown in Fig.~\ref{fig:DynamicArray}~{(a)}. We can push elements into a dynamic array, e.g., function \texttt{foo3()} in Fig.~\ref{fig:ArrayMapExample}~{(a)} pushes two integers, $10$ and $11$, into the dynamic array \texttt{a}. Now, the unique characteristic of dynamic arrays is that the location to store the pushed element is decided by a hash function, denoted by \texttt{HASH}. The first element will be stored in storage slot $t$ where $t = \mathtt{HASH(bytes32(}p))$ and $\mathtt{bytes32(}p)$ is the base slot number of the dynamic array padded into $32$~bytes long. The second element will be stored in slot $t+1$, and so on. Fig.~\ref{fig:DynamicArray}~{(b)} shows the locations to store the two elements pushed into \texttt{a}. Notice that the value of the base slot is updated to $2$ because \texttt{a} now has two elements. The rule for evaluating dynamic arrays can be obtained by the following rule. We use rule \texttt{$\texttt{Type}_\sigma$ exp} to check that the current expression is indeed a dynamic array. First, we obtain the l-value of the base expression to calculate the address where the expression is allocated. Second, we access to its value to check that the R-Value of the index expression is within the number of elements allocated in the dynamic array. With this information we calculate the final address through the Hashing function on the base address and the index accessed. 

\[
\footnotesize
\infer[\mbox{E-D-ARRAY}]{E(\sigma, \mathsf{Exp_b[Exp_i]}) \longrightarrow \langle \widehat{\sigma}', a \rangle}{
	\begin{array}{l}		
	\mathsf{Type_\sigma\ Exp_b = Type []} \\
	\mathsf{[{addr_b}]_{(ST\ Exp_b)_{\widehat{\sigma}}}^{Size (Type_{\widehat{\sigma}}\ Exp_b)} =} v \\
	i\leq (v-1)
	\end{array}
	& \begin{array}{l}
	E(\widehat{\sigma}, \mathsf{Exp_b}) \longrightarrow \langle \widehat{\sigma}', \mathsf{{addr_b}} \rangle \\
	E_R(\sigma, \mathsf{Exp_i}) \longrightarrow \langle \widehat{\sigma}, i \rangle \\
	a = \mathtt{addr32}(\mathtt{HASH}(\mathtt{bytes32}(\mathsf{addr_b})) + i)
	\end{array}
}
\]

Fig~\ref{fig:ArrayMapExample}~{(b)} shows another contract using a mapping \texttt{m}, which maps an unsigned integer to another. After it is declared, one storage slot is allocated to it as the base slot, but nothing is stored there, as shown in Fig.~\ref{fig:Mapping}~{(a)}. We can add key/value pair into a mapping, e.g., function \texttt{foo4()} in Fig.~\ref{fig:ArrayMapExample}~{(b)} adds two key/value pairs $(100:10)$ and $(200:11)$, where $100$ and $200$ are keys and $10$ and $11$ are their corresponding values. The unique characteristic of mappings is that the location to store values is decided by the hash function as well. For a key/value pair $(k:v)$ of a mapping with its base slot at $p$, the value $v$ will be stored in storage slot $t = \mathtt{HASH(bytes32(} p) \cdot \mathtt{bytes32(} k))$ where $\mathtt{bytes32(}k)$ is the key padded into $32$~bytes long and $\cdot$ is the concatenation operator. Fig.~\ref{fig:Mapping}~{(b)} shows the locations to store the two values $10$ and $11$. The rule for evaluating mappings can be performed by the following rule:

\[
\footnotesize
\infer[\mbox{E-MAPPING}]{E(\sigma, \mathsf{Exp_b[Exp_i]}) \longrightarrow \langle  \widehat{\sigma}', v \rangle}{
	\begin {array}{l}
	  v = \mathtt{addr32}(\mathtt{HASH}(\mathtt{bytes32}( \mathsf{{addr_b}}) \cdot \mathtt{bytes32}(i))) \\
	\begin{array}{ll}		
	\mathsf{Type_\sigma\ Exp_b = Mapping\ K\ T} \quad & 
	E( \widehat{\sigma}, \mathsf{Exp_b}) \longrightarrow \langle  \widehat{\sigma}', \mathsf{{addr_b}} \rangle \\
	\mathsf{Type_\sigma\ Exp_i = K} &
	E_R(\sigma, \mathsf{Exp_i}) \longrightarrow \langle  \widehat{\sigma}, i \rangle
	\end{array}		
	\end{array}
}
\]

The semantics of an internal function call is captured by the rules I-FUN and E-FUN to model when a function's role is an instruction or an expression. Every internal function call have its own name and type space to store local variables, arguments, and return values. Thus, a fresh name \texttt{N'} and type $\tau$ spaces are pushed into the stack $\amalg$, and the internal function call $\mathsf{id(Exp_1, \ldots, Exp_n)}$ is rewritten into a sequence of memory variable declaration statements to link arguments with function parameters, followed by the returning value, if any, and the function body $\mathsf{Block}$. When behaving as an expression the function call will return the L-Value of the returned variable, which value is added by the execution of the return instruction. The evaluation of both the instruction and the expression removes the name and type spaces from the stack the memory contains.

\[
\footnotesize
\infer[\mbox{I-FUN}]{\langle \sigma, \mathsf{id(Exp_1, \ldots, Exp_n) }  \rangle \longrightarrow \langle \widehat{\sigma}',. \rangle}{%
	\begin{array}{l}	
	\mathsf{\langle M_{\sigma'}.Push(N',\tau'), T_1 \, M \,\,p_1 = Exp_1; \ldots; T_n \, M \,\, p_n = Exp_n;\, Block}\rangle \rightarrow \langle \widehat{\sigma},. \rangle   \\
	\Gamma_p \mathsf{ id = Block } \quad \Gamma_t\mathsf{ id = ((T_1, \ldots, T_n)},\_) \quad M_{\widehat{\sigma}'}.\mathsf{Pop()}
	\end{array}
}
\]

\[
\footnotesize
\infer[\mbox{E-FUN}]{E(\sigma, \mathsf{id(Exp_1, \ldots, Exp_n) }) \longrightarrow \langle \widehat{\sigma}',v \rangle}{%
	\begin{array}{l}	
	\mathsf{\langle M_{\sigma'}.Push(N',\tau'), T_1 \,\, M\,\,p_1 = Exp_1; \ldots; T_n \,\, M\,\, p_n = Exp_n;\, T_r \,\, M\,\, p_r;\, Block}\rangle \rightarrow \langle \widehat{\sigma},. \rangle   \\	
	\mathsf{v = N_{\widehat{\sigma}} p_r \quad 
		\Gamma_p \mathsf{id = Block } \quad \Gamma_t\mathsf{ id = ((T_1, \ldots, T_n),(Type_r))} \quad M_{\widehat{\sigma}'}.\mathsf{Pop()}}
	\end{array}
}
\]

The rules for executing the statements in the function body are the similar to those used in high level programming languages and can be found in Appendix~\ref{subsec:StatementRule}. Here, we only highlight some important rules. For the ($\mathtt{return} \,\,\,\mathsf{Exp};$) statement, we obtain the R-Value of the returning expression, and we assign it to the return variable declared when calling the function. Therefore the returning value is available to the caller after returning from the call when the function is in an expression.


\[
\footnotesize
\infer[\mbox{RETURN}]{\langle \sigma, \mathtt{return} \,\,\,\mathsf{Exp} \,; \rangle \longrightarrow \langle \widehat{\sigma'}, . \rangle}{%
	\begin{array}{l}		
	E_R(\sigma, \mathsf{Exp}) \longrightarrow \langle \widehat{\sigma}, v \rangle 
	\end{array}
	& \begin{array}{l}		
	\mathsf{[{N}_{\widehat{\sigma}} p_r]_{(ST\ exp)_{\widehat{\sigma}}}^{Size\ ({\tau}_{\widehat{\sigma}} p_r)} =} v
	\end{array}
}
\]

The semantics of external function calls is captured by the following two rules. The E-FUN$_1$ rule is applicable when one contract instance wants to call an external function $\mathsf{id_2}$ of another contract instance located in address $\alpha$. After looking up the configuration of the callee contract, $\ddot{\sigma}$, we push the caller's configuration $\sigma$ into the callee's configuration stack such that when the external function call is finished, the caller's configuration can be restored. Then, the external function call is translated into an internal function call under the callee's configuration. The E-FUN$_2$ rule is applicable when one contract instance just wants to send ether to another contract instance located in address $\alpha$ without calling any function. In this case, the fallback function of the callee contract will be invoked.

\[
\footnotesize
\infer[\mbox{E-FUN}_1]{\langle \sigma, \mathsf{id_1.id_2(Exp_1, \ldots, Exp_n)\mathtt{.value}\mathsf{(Exp_{eth})}.\mathtt{gas}\mathsf{(Exp_{gas})}}  \rangle \longrightarrow \langle \ddot{\sigma}', \mathsf{id_2(Exp_1, \ldots, Exp_n)} \rangle}{%
	\begin{array}{l}
		\mathsf{id_1: Contract} \mbox{ in address } \alpha \\
		\mathsf{id_2: } \,\,\mathtt{function} \,\, \mathsf{(Type_1, \ldots, Type_n )}
	\end{array}
	& \begin{array}{l}
		\Delta(\alpha) = \ddot{\sigma} \\
		\ddot{\Omega}' = \ddot{\Omega}.\mathtt{push}(\sigma)
	  \end{array}
	& \begin{array}{l}
		E(\sigma, \mathsf{Exp_{eth}}) \longrightarrow \langle \sigma, m \rangle \\
		E(\sigma, \mathsf{Exp_{gas}}) \longrightarrow \langle \sigma, n \rangle \\
	\end{array}
}
\]

\[
\footnotesize
\infer[\mbox{E-FUN}_2]{\langle \sigma, \mathsf{id}\mathtt{.call.value}\mathsf{(Exp_{eth})\mathtt{.gas}\mathsf{(Exp_{gas})}}  \rangle \longrightarrow \langle \ddot{\sigma}', \mathtt{fallBack()} \rangle}{%
	\begin{array}{l}
		\mathsf{id: Contract} \mbox{ in address } \alpha
	\end{array}
	& \begin{array}{l}
	     	E(\sigma, \mathsf{Exp_{eth}}) \longrightarrow \langle \sigma, m \rangle \\
	     	E(\sigma, \mathsf{Exp_{gas}}) \longrightarrow \langle \sigma, n \rangle
	       \end{array}
	& \begin{array}{l}
		\Delta(\alpha) = \ddot{\sigma}\\
		\ddot{\Omega}' = \ddot{\Omega}.\mathtt{push}(\sigma)
	  \end{array}
}
\]

\section{Solidity Semantics in K-framework} \label{sec:SemanticsInK}

In this section, we introduce the Solidity semantics we have implemented in K-framework\cite{K} by illustrating some important rules.
This implementation reflects the idea of the formal specification we introduce in section \ref{sec:SoliditySemantics} and involves over 200 rules. The K definition of the Solidity semantics takes up more than 2000 lines and consists of three main parts, namely syntax, configuration and rules. The syntax can be found in \cite{Sol}, and the configuration of the semantics is attached in the appendix \ref{solidity-configuration}, so we do not explain these parts in detail when presenting the rules. Due to limit of space, we only show some important rules here.

\Newrule{Elementary-TypeName}{
\RWrule{
\code{pcsContractPart(C:Id, X:ElementaryTypeName Y:Specifiers Z:Id = E;) }
}{
\code{.K}
}{k}\\

\RruleS{
\RruleM{\code{C}}{cname}
\RruleM{\code{N:Int$\Rightarrow$N +Int 1}}{vnum}
\RWrule{
\code{.Map}
}{
\code{N $\mapsto$ Z}
}{vId}
\\
\RWrule{
\code{.Map}
}{
\code{Z $\mapsto$ !Num:Int}
}{variables}

\RWrule{
\code{.Map}
}{
\code{Z $\mapsto$ X}
}{typename}\\
\RWrule{
\code{.Map}
}{
\code{!Num $\mapsto$ E}
}{cstore}
}
{contract}
}

Let us start with the state variable declaration for elementary type names (shown in Rule \rulename{Elementary-TypeName}).
When there is a state variable declaration for elementary type names in contract parts, we take a record of it in the $contract$ cell. The number of variables will be increased by one (In $vnum$ cell). 
The symbol \codec{!Num} means generating a fresh integer number as the address of the variable. The two pairs: (1) \codec{Z} to its address \codec{!Num} and (2) the address \codec{!Num} to its value \codec{E} are added to the $variables$ and $cstore$ cells, respectively.

\Newrule{Function-Definition}{
\RWrule{
\code{pcsContractPart(C:Id, function F:Id (Ps1:Parameters)}\\ \code{FQ:FunQuantifiers returns  (Ps2:Parameters)$\{$B$\}$) }
}{
\code{.K}
}{k}\\

\RruleM{\code{C}}{cname}
\RWrule{
\code{.Map}
}{
\code{F $\mapsto$ CF}
}{cfunction}
\\
\RruleM{\code{CF:Int$\Rightarrow$CF +Int 1 }}{cntfunction}
\\
\code{.Bag}$\Rightarrow$
\RruleS{
\RruleM{\code{CF}}{fId}
\RruleM{\code{Ps1}}{inputparameters}
\RruleM{\code{Ps2}}{outputparameters}\\
\RruleM{\code{FQ}}{FunQuantifiers}
\RruleM{\code{true}}{FunCon}
\\
\RruleM{\code{0}}{return}
\RruleM{\code{B}}{body}
}
{function}
}

Rule \rulename{Function-Definition} is how we deal with function definitions. A mapping from the function name to function Id is generated in the cell \emph{cfunction}, which enables us to identify each function by using function Id. A bag of the cell \emph{function} containing function Id, input and output parameters, function quantifiers, function body, etc, is created for this function definition. In this way, we can retrieve the details of this function in the cell \emph{function}.

\Newrule{Internal-Function-Call}{
\RWrule{
\code{functionCall(F:Id ; Es:Values)}
}{
\code{FunQs(FQ,F)$\curvearrowright$ Call(F,Es)}
}{k}\\

\RruleS{\code{ListItem(CI:Int)}}{contractStack}\\
\RruleS{\RruleM{\code{CI}}{ctId}
\RruleM{\code{Cn:Id}}{ctname}}{contractInstance}\\
\RruleS{
\RruleS{
\RruleM{\code{Cn}}{cname}\Rrule{
\code{F$\mapsto$CT:Int}
}{cfunction}}{contract}
}{contracts}\\

\RruleS{
\RruleM{\code{CT}}{fId}
\RruleM{\code{Ps}}{inputparameters}
\RruleM{\code{FQ}}{FunQuantifiers}\\
\RruleM{\code{Con}}{FunCon}
\RruleM{\code{B}}{body}
}
{function}
}

As for internal function call (shown in Rule \rulename{Internal-Function-Call}), we need to process the function quantifiers first, and then \emph{Call} which deals with the execution of the function body. The function Id is obtained from the Id of current contract instance and the name of the contract defining the function, to retrieve the details of this function.

\Newrule{Call}{
\RWrule{
\code{Call(F:Id,Es:Values)}
}{
\code{ BindParam(Ps,Es)$\curvearrowright$ if(Con){B}}
}{k}\\

\RruleS{\code{ListItem(CI:Int)}}{contractStack}\\
\RruleS{\RruleM{\code{CI}}{ctId}
\RruleM{\code{Cn:Id}}{ctname}}{contractInstance}\\
\RruleS{
\RruleS{
\RruleM{\code{Cn}}{cname}\Rrule{
\code{F$\mapsto$CT:Int}
}{cfunction}}{contract}
}{contracts}\\

\RruleS{
\RruleM{\code{CT}}{fId}
\RruleM{\code{Ps}}{inputparameters}
\RruleM{\code{FQ}}{FunQuantifiers}\\
\RruleM{\code{Con}}{FunCon}
\RruleM{\code{B}}{body}
}
{function}
}

Rule \rulename{Call} deals with the execution of function body. It first binds the parameters of function call in current execution environment. After that, the body of the function is executed with a condition specified by the function quantifiers. If there is no modifier invocation in the function quantifiers, the condition is always true.

\Newrule{External-Function-Call}{
\RWrule{
\code{functionCall(C:Int ; F:Id ; Es:Values ; M:Msg)}
}{
\code{createTransaction(L)$\curvearrowright$ functionCall(F ; Es)$\curvearrowright$returnContext(C)}
}{k}\\

\RruleM{\code{(.List $\Rightarrow$ ListItem(C)) L:List ListItem(-1)}}{contractStack}

\RruleM{\code{ M1$\Rightarrow$M}}{Msg}
\RruleS{\code{ (.List $\Rightarrow$ ListItem(M1))}}{MsgStack}\\
\RruleS{\code{ (.List => ListItem(F))}}{functionStack}
}

Rule \rulename{External function call} is associated with transactions. The input parameters of \emph{External function call} are the Id of the contract instance to be called, the name and parameters of the function and \codec{Msg} which contains information about this transaction.  The number of transactions that have been executed is counted in \codec{createTransaction(L)}, followed by an internal function call and context return. Meanwhile, the Id of contract instance, current \codec{Msg} and the function to be called are stored in the corresponding stacks, and the cell \codec{Msg} is updated.

\section{Evaluation} \label{sec:Evaluation}

We evaluate the proposed Solidity semantics from two perspectives: the first one is its coverage, and the second is the ability to detect vulnerabilities in smart contracts. Our test set is obtained from ~\cite{Sol}. In Section~\ref{subsec:Coverage}, we show that the proposed Solidity semantics covers most of the important semantics specified by the official Solidity document~\cite{Sol} and is consistent with the official Solidity compiler~\cite{remix}. In Section~\ref{subsec:Verification}, we show that some variants of DAO attacks can be detected by using the proposed semantics, which facilitates the verification of smart contracts.

\subsection{Coverage and Testing} \label{subsec:Coverage}

We evaluate and test the proposed Solidity semantics by using the official Solidity compiler Remix~\cite{remix}. The evaluation is done by manually comparing the results of our implementation in K-framework with the results of the Remix compiler. We consider the proposed semantics is correct if the result is consistent with that of the Remix compiler. We list the coverage of our Solidity semantics in Table~\ref{T1} from a variety of perspectives specified by the syntax provided by the official Solidity document\cite{Sol}.

\begin{table}[tb]
\scriptsize
\sf
\centering
\caption{Coverage of The Proposed Solidity Semantics\label{T1}}
\begin{tabular}{l@{\hspace{12mm}}c@{\hspace{12mm}}l@{\hspace{12mm}}c}
\toprule
\textbf{Perspectives} &\textbf{Coverage} & \textbf{Perspectives} &\textbf{Coverage}\\
\midrule
\textbf{Syntax}\\
\texttt{Basic Syntax} & FC&\textbf{Using For} & N\\
\texttt{Hex Number/Hex Literal} & N&\textbf{Event} & N\\
\texttt{Assembly} & N&\textbf{Inheritance} & N\\
\midrule
\textbf{Storage}&&\textbf{Statements} \\
\emph{Elementary TypeName} &&\texttt{If Statement} & FC \\
\quad \texttt{address} &FC&\texttt{While Statement} & FC \\
\quad\texttt{bool} &FC&\texttt{For Statement} & FC\\
\quad\texttt{string} &FC&\texttt{Block} & FC \\
\quad\texttt{var} &FC&\texttt{Inline Assembly} & N\\
\quad\texttt{int256} &FC &\texttt{Statement}\\
\quad\texttt{Other Int Size} & N&\texttt{Do While Statement} & FC \\
\quad\texttt{uint256} & FC&\texttt{Place Holder Statement} & FC\\
\quad\texttt{Other Uint Size} & N &\texttt{Continue} & N \\
\quad\texttt{Byte} & N&\texttt{Break} & N \\
\quad\texttt{Fixed} & N&\texttt{Return} & FC \\
\quad\texttt{Ufixed} & N&\texttt{Throw} & N \\
\emph{User Defined TypeName} & P&\texttt{Simple Statement} & FC \\
\emph{Mapping} & FC \\
\emph{Array TypeName} & FC \\
\emph{Function TypeName} & N \\
\midrule
\textbf{Functions}&&\textbf{Expressions}\\
\emph{Function Definition}&&\texttt{Bitwise Operations} & N \\
\quad\texttt{Constructor} & FC&\texttt{Other Expressions} & FC\\
\quad\texttt{Normal Functions} & FC\\
\quad\texttt{Fallback Functions} & FC\\
\quad\texttt{Modifier} & FC\\
\quad\texttt{StateMutability} & N\\
\quad\texttt{Specifier} & N\\
\emph{Function Call} \\
\quad\texttt{Internal Function Call} & FC \\
\quad\texttt{External Function Call} & FC \\
\bottomrule
\end{tabular}
\begin{tablenotes}
	\item \texttt{FC}: Fully Covered and Consistent with Solidity IDE
	\item \texttt{P}: Partially Covered and Consistent with Solidity IDE for Covered Parts
	\item \texttt{N}: Not Covered
\end{tablenotes}
\end{table}

%

From Table~\ref{T1}, we can observe that the proposed Solidity semantics covers most of the syntax except Hex number and literal, and Solidity assembly code. As for storage, our semantics implementation in K-framework covers the following elementary types: \texttt{address}, \texttt{bool}, \texttt{string}, \texttt{var}, \texttt{int256} and \texttt{uint256}. User-defined type is partially covered, including \texttt{struct} and  contract instances. Mappings and arrays are covered, while function types are not. In addition, most parts of semantics associated with functions are covered except state mutability and specifiers which are ignored in our current semantics implementation in the execution. Furthermore, a majority of statements and expressions are covered. For all the parts of covered semantics, they are considered to be correct since the execution behaviours involved are consistent with the official Solidity compiler.

Although our semantics implementation in K-framework is not complete yet, it covers most of the semantics in smart contracts. Actually, the set of semantics in which vulnerabilities of smart contracts lie has already been covered. Taking the DAO attack (c.f. Section~\ref{apx:DAO}) as an example, the two vulnerabilities, reentrancy and call to the unknown, are mainly associated with the semantics of function calls. For the uncovered parts, they can be either ignored or transformed into the semantics that are covered such that the missing semantics does not have a big impact on the execution behaviours. Thus, our implementation of Solidity semantics can be used in the verification of smart contracts.

\subsection{Detecting DAO Attacks} \label{subsec:Verification}
We briefly introduce the DAO attack in Appendix~\ref{apx:DAO}. Interested readers can get the detail there. We evaluate four variants of DAO attacks by using our implementation of Solidity semantics in K-framework. We simulate the behaviour of users on the blockchain by using a \texttt{Main} contract in which transactions are generated. Notice that the mining process on the blockchain is not modeled. The evaluation result shows that these DAO attacks can be fully executed, and the non-reentrant behaviour can be detected from the values of cells in the configuration. The former indicates that our implementation of Solidity semantics is executable, while the latter shows that some vulnerabilities in smart contracts can be detected with the executable semantics, contributing to the verification of security properties in smart contracts.


\section{Related Works} \label{sec:RelatedWorks}

K-framework (\kname)~\cite{rosu-serbanuta-2010-jlap} is a rewrite logic based  formal \emph{executable} semantics definition framework. The semantics of various programming languages
have been defined using \kname, such as Java \cite{bogdanas-rosu-2015-popl}, C \cite{hathhorn-ellison-rosu-2015-pldi,ellison-rosu-2012-popl}, and Javascript \cite{park-stefanescu-rosu-2015-pldi}. Particularly, the executable semantics of the EVM(Ethereum Virtual Machine), the bytecode language of smart contracts, has been created in K-framework\cite{KEVM}.
\kname\, backends, like the  Isabelle theory generator, model checker, and deductive verifier, can be utilized to prove properties on the semantics and construct verification tools.
For instance,
\kname\, provides pre- and post-condition verification by using
Matching Logic \cite{rosu-2017-lmcs}. Also, the Reachability Logic prover in \kname\ can be used to verify properties specified as reachability claims.
In fact, \kname\, aims to provide a semantics-based program verifier for all languages \cite{oopsla/StefanescuPYLR16}.

\section{Conclusion And Future Work} \label{sec:Conclusion}
In this paper, we introduce our executable operational semantics of Solidity in K-framework. We present an abstract model of semantics and illustrate some important rules implemented in K-framework. Experiment results show that our Solidity semantics has already covered most of the semantics specified by the official Solidity documentation\cite{Sol}, and the covered semantics are consistent with the official Solidity compiler\cite{remix}. Furthermore, we show that our semantics can be used to verify certain properties in smart contracts.


For future work, we plan to complete the Solidity semantics in K-framework to completely cover all the features os Solidity. Additionally, we plan to approach verification of attacks in Solidity contracts, identifying different kinds of vulnerabilities in smart contracts \cite{DBLP:conf/post/AtzeiBC17} and constructing verification properties against attacks.

\clearpage

\appendix


\section{Solidity Configuration}
\label{solidity-configuration}
{
\RruleM{
\RruleM{
\RruleM{
\RruleM{\code{\$PGM:SourceUnit}}{k}
\RruleM{\code{Map}}{env}\\
\RruleM{\code{ListItem(-1)}}{contractStack}\\
\RruleM{\code{List}}{functionStack}
\RruleM{\code{List}}{newStack}
}{thread*}
}{threads}\\
\RruleM{\code{Map}}{store}
\RruleM{\code{1:Int}}{vposition}\\
\RruleM{
\RruleM{\code{0:Int}}{contractnum}\\
\RruleS{
\RruleM{\code{K}}{cname}
\RruleM{\code{ (-1):Int}}{cId}
\RruleM{\code{Map}}{cfunction}\\
\RruleM{\code{0:Int}}{vnum}
\RruleM{\code{Map}}{vId}
\RruleM{\code{Map}}{variables}\\
\RruleM{\code{Map}}{typename}
\RruleM{\code{Map}}{cstore}
\RruleM{\code{Map}}{dtype}\\
\RruleM{\code{Map}}{dstore}
\RruleM{\code{1:Int}}{dim}
\RruleM{\code{Map}}{vdim}
}
{contract*}
}{contracts}\\
\RruleM{
\RruleM{
\RruleM{\code{0:Int}}{fId}
\RruleM{\code{Parameters}}{inputparameters}\\
\RruleM{\code{Parameters}}{outputparameters}
\RruleM{\code{K}}{body}\\
\RruleM{\code{K}}{FunQuantifiers}
\RruleM{\code{K}}{return}
\RruleM{\code{true}}{FunCon}
}{function*}\\
\RruleM{\code{0:Int}}{cntfunction}\\
}{functions}\\
\RruleM{
\RruleS{
\RruleM{\code{(-1):Int}}{ctId}
\RruleM{\code{K}}{ctname}\\
\RruleM{\code{0:Int}}{ctnum}
\RruleM{\code{Map}}{ctvId}\\
\RruleM{\code{Map}}{ctContext}
\RruleM{\code{Map}}{ctType}\\
\RruleM{\code{Map}}{ctstore}
\RruleM{\code{Map}}{memory}
}{contractInstance*}\\
\RruleM{\code{0:Int}}{cntContracts}\\
}{contractInstances}\\
\RruleM{
\RruleM{\code{1:Int}}{cnttran}
\RruleM{\code{0 |-> "Main"}}{trancomputation}
}{transactions}\\
\RruleM{\code{K}}{Msg}
\RruleM{\code{List}}{MsgStack}
}{T}
}

\section{DAO Attack} \label{apx:DAO}

DAO \cite{DAO} is a contract which implements a platform for crowd-funding. As reported before, $\$$ 60M can be taken under the control of the attacker in the DAO attack \cite{Attack}, which has a huge impact in the financial aspects. A simplified version of DAO attacks is shown in Fig.~\ref{fig:DAOAttack}. The smart contract \texttt{Bank} is used to collect funding from different clients. A client can invoke the function \texttt{deposit()} to deposit ETH to its account in the \texttt{Bank} contract, or invoke the function \texttt{withdraw()} to withdraw his/her credit. The malicious contract \texttt{Attack} can be used to stole ETH from the contract \texttt{Bank}. Let us assume that the contract \texttt{Bank} has accumulated a certain amount of ETH. 

The attack can be launched in the following procedures.  First, the contract \texttt{Attack} is created and deployed with its state variable \texttt{target} pointing to the victim bank in its constructor. After that, the function \texttt{addToBalance()} is invoked by the attacker to deposit $2$~wei\footnote{Wei is the minimum unit of ETH. $1$~wei $= 10^{-18}$ ETH.} to the contract \texttt{Bank}. As a result, the balance of \texttt{Bank} is increased by 2 wei. Since the contract \texttt{Attack} is just created and deployed on the blockchain, its initial credit in \texttt{Bank} is 0. After deposit, the credit of the sender in this case, should become 2 wei. Subsequently, the attacker invokes the function \texttt{withdrawBalance()} of contract \texttt{Attack} to withdraw 2 wei from the \texttt{Bank} contract. In the function \texttt{withdraw()} of contract \texttt{Bank}, the amount to be withdrawn is sent to contract \texttt{Attack} first (line~$15$), and then the amount is deduced from the credit of contract \texttt{Attack} (line~$16$). When contract \texttt{Attack} receives the withdrawn amount (due to line~$15$ of \texttt{Bank}), its fallback function (lines~$16$--$18$) is invoked. Inside the function body of its fallback function, it maliciously invokes the \texttt{withdraw()} function of \texttt{Bank} again. At this point, the amount to be withdrawn has not been deduced from \texttt{Attack}'s credit (line~$16$ of \texttt{Bank}), which makes the condition checking in line $14$ of \texttt{Bank} still valid. Thus, contract \texttt{Attack} is able to withdraw money from contract \texttt{Bank} recursively until the balance of \texttt{Bank} becomes zero.

The vulnerability comes from the fact that the \texttt{withdraw()} function of contract \texttt{Bank} is not reentrant due to the wrong order of lines~$15$ and $16$. The amount to be withdrawn should be deduced from the credit first and then sent to the withdrawer. If we switch the order of lines~$15$ and $16$ of contract \texttt{Bank}, then the function \texttt{withdraw()} becomes reentrant.


\begin{figure}[tb]
\centering
\begin{minipage}{0.5\linewidth}
\scriptsize
\centering
\begin{lstlisting}[frame=single,numbers=left,numberstyle=\tiny]
contract Bank {
 mapping(address=>uint) credit;

 function getUserBalance(address user)
   constant returns(uint) {
   return credit[user];
 }

 function deposit() payable{
   credit[msg.sender] += msg.value;
 }

 function withdraw(uint amount){
   if(credit[msg.sender] >= amount){
     msg.sender.call.value(amount)();
     credit[msg.sender] -= amount;
   }
 }
}	
\end{lstlisting}
(a) The \texttt{Bank} Smart Contract
\end{minipage}
~ \hspace{2mm} ~
\begin{minipage}{0.4\linewidth}
\scriptsize
\centering
\begin{lstlisting}[frame=single,numbers=left,numberstyle=\tiny]
contract Attack {
 Bank target;

 function Attack(address addr){
   target = Bank(addr);
 }

 function addToBalance(){
   target.deposit.value(2)();
 }

 function withdrawBalance(){
   target.withdraw(2);
 }

 function() payable{
   target.withdraw(2);
 }
}
\end{lstlisting}
(b) The \texttt{Attack} Smart Contract
\end{minipage}
\caption{DAO Attack} \label{fig:DAOAttack}
\end{figure}

\section{Rules of Statements} \label{subsec:StatementRule}

\[
\footnotesize
\begin{array}{lcr}
\infer[\mbox{VD}_1]{\langle \sigma, \mathsf{Type\,\,\, id = Exp \,;} \rangle \longrightarrow \langle \widehat{\sigma}', \cdot \rangle }{%
	\begin{array}{l}		
	E_R(\sigma, \mathsf{Exp}) \longrightarrow \langle \widehat{\sigma}, v \rangle \\
	\mathsf{N_{\widehat{\sigma}'} id = \ceil*{\Lambda_{\widehat{\sigma}}}^{(l, Type)}} \\ 
	\mathsf{{\Lambda}_{\widehat{\sigma}'} = \Lambda_{\widehat{\sigma}}\uparrow(l, Type)}
	\end{array}
	& \begin{array}{l}	  
	\mathsf{\tau_{\widehat{\sigma}'} id = Type} \\	
	\mathsf{id} \not\in N_{\widehat{\sigma}} \\	
	\mathsf{[N_{\widehat{\sigma}'} id]_{\widehat{\sigma}'}^{Size\ Type} =} v
	\end{array}}
&\hspace{5mm} &
\infer[\mbox{VD}_2]{\langle \sigma, \mathsf{Type\,\,\, id = Exp \,;} \rangle \longrightarrow \langle \widehat{\sigma}', \cdot \rangle }{%
	\begin{array}{l}		
	E_R(\sigma, \mathsf{Exp}) \longrightarrow \langle \widehat{\sigma}, v \rangle \\	 
	\mathsf{M_{\widehat{\sigma}'} = fr_{\widehat{\sigma}}(id,Type,}v)
	\end{array}
	& \begin{array}{l}
	\mathsf{id} \not\in N_{\widehat{\sigma}}
	\end{array}
}
\end{array}
\]

\[
\footnotesize
\infer[\mbox{I-FUN}]{\langle \sigma, \mathsf{id(Exp_1, \ldots, Exp_n) }  \rangle \longrightarrow \langle \widehat{\sigma}',. \rangle}{%
	\begin{array}{l}	
	\mathsf{\langle M_{\sigma'}.Push(N',\tau'), T_1 \, M \,\,p_1 = Exp_1; \ldots; T_n \, M \,\, p_n = Exp_n;\, Block}\rangle \rightarrow \langle \widehat{\sigma},. \rangle   \\
	\Gamma_p \mathsf{ id = Block } \quad \Gamma_t\mathsf{ id = ((T_1, \ldots, T_n)},\_) \quad M_{\widehat{\sigma}'}.\mathsf{Pop()}
	\end{array}
}
\]

\[
\footnotesize
\infer[\mbox{ASSIGN}]{\langle \sigma, \mathsf{Exp}_1 = \mathsf{Exp}_2 \,; \rangle \longrightarrow \langle \sigma', \cdot \rangle}{%
	\begin{array}{l}
	E_R(\sigma, \mathsf{Exp_2}) \longrightarrow \langle \widehat{\sigma}, v \rangle \\    
	E(\widehat{\sigma}, Exp1) \longrightarrow (\widehat{\sigma'}, addr) \\
	\mathsf{[addr]_{(ST\ Exp_2)_{\widehat{\sigma}'}}^{Size (Type_{\widehat{\sigma}}\ Exp_2)} =} v 
	\end{array}
}
\]

\[
\footnotesize
\infer[\mbox{SEQ}]{\langle \sigma, \mathsf{Stmt}_1 \, \mathsf{Stmt}_2 \rangle \longrightarrow \langle \sigma'', \cdot \rangle }{%
	\langle \sigma, \mathsf{Stmt}_1 \rangle \longrightarrow \langle \sigma', \cdot \rangle
	& \langle \sigma', \mathsf{Stmt}_2 \rangle \longrightarrow \langle \sigma'', \cdot \rangle
}
\]

\[
\footnotesize
\infer[\mbox{COND}_1]{\langle \sigma, \mathtt{if} \, \mathsf{(Exp) \, \{} \, \mathsf{Block}_1 \, \} \, \mathtt{else \, \{ } \, \mathsf{Block}_2 \, \} \rangle \longrightarrow \langle \widehat{\sigma}', \cdot \rangle }{%
	E_R(\sigma, \mathsf{Exp}) \longrightarrow \langle \widehat{\sigma}, \top \rangle \\  
	& \langle \widehat{\sigma}, \mathsf{Block}_1 \rangle \longrightarrow \langle \widehat{\sigma}', \cdot \rangle
}
\]

\[
\footnotesize
\infer[\mbox{COND}_2]{\langle \sigma, \mathtt{if} \, \mathsf{(Exp) \, \{} \, \mathsf{Block}_1 \, \} \, \mathtt{else \, \{ } \, \mathsf{Block}_2 \, \} \rangle \longrightarrow \langle \widehat{\sigma}', \cdot \rangle }{%
E_R(\sigma, \mathsf{Exp}) \longrightarrow \langle \widehat{\sigma}, \bot \rangle \\  
	& \langle \sigma, \mathsf{Block}_2 \rangle \longrightarrow \langle \widehat{\sigma}', \cdot \rangle
}
\]

\[
\footnotesize
\infer[\mbox{WHILE}_1]{\langle \sigma, \mathtt{while} \, \mathsf{(Exp) \, \{} \, \mathsf{Block} \, \} \rangle \longrightarrow \langle \widehat{\sigma}', \cdot \rangle }{%
	E_R(\sigma, \mathsf{Exp}) \longrightarrow \langle \widehat{\sigma}, \bot \rangle
}
\]

\[
\footnotesize
\infer[\mbox{WHILE}_2]{\langle \sigma, \mathtt{while} \, \mathsf{(Exp) \, \{} \, \mathsf{Block} \, \} \rangle \longrightarrow \langle \widehat{\sigma}', \mathtt{while} \, \mathsf{(Exp) \, \{} \, \mathsf{Block} \, \} \rangle }{%
    E_R(\sigma, \mathsf{Exp}) \longrightarrow \langle \widehat{\sigma}, \top \rangle \\ 
	& \langle \widehat{\sigma}, \mathsf{Block} \rangle \longrightarrow \langle \widehat{\sigma}', \cdot \rangle
}
\]

\[
\footnotesize
\begin{array}{lcl}
\infer[\mbox{Skip}_1]{\langle \sigma, \mathsf{Exp \,;}  \rangle \longrightarrow \langle \sigma', \cdot \rangle}{
	\Omega \mbox{ is empty}
}
&
\hspace{10mm}
&
\infer[\mbox{Skip}_2]{\langle \sigma, \mathsf{Exp \,;}  \rangle \longrightarrow \langle \sigma'', \mathsf{Exp \,;} \rangle}{
	\Omega \mbox{ is not empty}
	& \sigma'' = \Omega.\mathtt{top()}
	& \Omega.\mathtt{pop()}
}
\end{array}
\]

\section{Rules of Evaluations}\label{subsec:MVeval}
{\footnotesize
	
	\begin{tabular}{ccc}		
		\infer[\textsc{SR1}]{\mathsf{SizeR\ n \langle \rangle = n}}
		{ 
			\begin{tabular}{l}
				$ $
			\end{tabular}
		}
		\hspace{0.2cm} &		
		\infer[\textsc{SR2}]{\mathsf{SizeR\ n\ \langle uint_{m}@Tl \rangle = m}}
		{ 
			\begin{tabular}{l}
				$\mathsf{SizeR\ (\ceil*{n}^{(l,uint_{m})}+(Size\ uint_{m}))\ \langle Tl \rangle = m}$
			\end{tabular}
		}
		\hspace{0.2cm} &
		\infer[\textsc{SR3}]{\mathsf{SizeR\ n\ \langle T@Tl \rangle = m}}
		{ 
			\begin{tabular}{l}
				$\mathsf{SizeR\ (\ceil*{n}^l+(Size\ T))\ \langle Tl \rangle = m}$
			\end{tabular}
		}
	\end{tabular}
	\vspace{0.3cm}

	\begin{tabular}{ccc}

		\infer[\textsc{Size1}]{\mathsf{Size\ uint_{m}  = 2^{m-3}}}
		{
			\begin{tabular}{l}
				$ $ 
			\end{tabular}
		}
		\hspace{0.2cm}&
		\infer[\textsc{Size2}]{\mathsf{Size\  T[n] = \ceil*{( n*(Size\ T))}^l}}
		{ 
			\begin{tabular}{l}
				$ $
			\end{tabular}
		}
		\hspace{0.2cm}&					
		\infer[\textsc{Size3}]{\mathsf{Size\ \langle T@Tt \rangle = \ceil*{n}^l}}
		{
			\begin{tabular}{l}
				$\mathsf{SizeR\ 0\  T@Tt = n}$ 
			\end{tabular}
		}		
	\end{tabular}  
	\vspace{0.3cm}
	
	\begin{tabular}{cccc}

		\infer[\textsc{Size4}]{\mathsf{Size\ T[]  = l}}
		{
			\begin{tabular}{l}
			\end{tabular}
		}
		\hspace{0.2cm}&
		\infer[\textsc{Size5}]{\mathsf{Size\ map\ K\ T = l}}
		{ 
			\begin{tabular}{l}
				$ $
			\end{tabular}
		}
		\hspace{0.2cm}&
		\infer[\textsc{Size6}]{\mathsf{Size\ Call\ fid() = Size\ T}}
		{ 
			\begin{tabular}{l}
				$\mathsf{\Gamma_{type} fid = T}$
			\end{tabular}
		}  
     	\hspace{0.2cm}&
		\infer[\textsc{Size7}]{\mathsf{Size\ T\ ref = l}}
		{ 
			\begin{tabular}{l}
				$ $
			\end{tabular}
		}  
	\end{tabular}

	\begin{tabular}{ccc}

		\infer[\textsc{Type1}]{\mathsf{Type_\sigma\ exp[exp_i]  =T}}
		{
			\begin{tabular}{l}
				$\mathsf{Type_\sigma\ exp = T[\_]}$\\
				$\mathsf{Type_\sigma\ exp_i = uint_m}$
			\end{tabular}
		}
		\hspace{0.2cm}&
		\infer[\textsc{Type2}]{\mathsf{Type_\sigma\ exp.k = T_k}}
		{ 
			\begin{tabular}{l}
				$\mathsf{k \leq i} \quad
				\mathsf{Type\ exp = \langle T_0 \cdots T_i  \rangle }$ \\	           
			\end{tabular}
		}
		\hspace{0.2cm}&					
		\infer[\textsc{Type3}]{\mathsf{Type_\sigma\ id = \sigma_t v}}
		{
			\begin{tabular}{l}
				$ $
			\end{tabular}
		}
	\end{tabular}  
	\vspace{0.3cm}
	
	\begin{tabular}{ccc}
		\infer[\textsc{Type4}]{\mathsf{Type_\sigma\ map_{acc}\ exp\ exp_i  = \mathbb{T}}}
		{
			\begin{tabular}{l}
				$\mathsf{Type_\sigma\ exp = map\ K\ T}$\\
				$\mathsf{Type_\sigma\ exp_i = K}$
			\end{tabular}
		}
		\hspace{0.2cm}&
		\infer[\textsc{Type5}]{\mathsf{Type_\sigma\ Call\ fid() = T}}
		{ 
			\begin{tabular}{l}
				$\mathsf{\Gamma_{type} fid = T}$
			\end{tabular}
		}
	\end{tabular}  
	\vspace{0.3cm}
	
	\begin{tabular}{ccc}
		\infer[\textsc{Type6}]{\mathsf{Type_\sigma\ map_{acc}\ exp\ exp_i  = \mathbb{T}}}
		{
			\begin{tabular}{l}
				$\mathsf{Type_\sigma\ exp = ref\ map\ K\ T}$\\
				$\mathsf{Type_\sigma\ exp_i = K}$
			\end{tabular}
		}
		\hspace{0.2cm}&
			\infer[\textsc{Type7}]{\mathsf{Type_\sigma\ exp[exp_i]  =T}}
		{
			\begin{tabular}{l}
				$\mathsf{Type_\sigma\ exp = ref\ T[\_]}$\\
				$\mathsf{Type_\sigma\ exp_i = uint_m}$
			\end{tabular}
		}
		\hspace{0.2cm}&
		\infer[\textsc{Type8}]{\mathsf{Type_\sigma\ exp.k = T_k}}
		{ 
			\begin{tabular}{l}
				$\mathsf{k \leq i} \quad
				\mathsf{Type\ exp = ref\ \langle T_0 \cdots T_i  \rangle }$ \\	           
			\end{tabular}
		}
	\end{tabular}  
	\vspace{0.3cm}

	\begin{tabular}{ll}
		&$\mathsf{\ceil*{addr}^{(l,PrimType)} = if\ addr + Size\ PrimType \leq l\ then\ addr\ else\ ((addr\ MOD\ l)+1)*l }$ \\
		$|$ & $\mathsf{ \ceil*{addr}^{(l,Type)} = if\  addr\ MOD\ l = 0\ then\ addr\ else\  ((addr\ MOD\ l)+1)*l}$  \\	
		& $\mathsf{addr \uparrow (l,Type) = \ceil*{addr}^{(l,Type)} + Size\ Type}$ 
	\end{tabular}

	Accessing bytes memory  :
	$\mathsf{[addr]_{m}^s = v} \leftrightarrow \forall i. 0 \leq i < s \rightarrow m (addr + i) = v\ i $
	
	\[
	\footnotesize
	\begin{array}{lcr}
	\infer[\mbox{E-ID1}]{E(\sigma, \mathsf{id}) \longrightarrow \langle \sigma,  (N_{\Psi_{\sigma}} \mathsf{id}) ) \rangle}{%
		\mathsf{id} \in N_{\Psi_{\sigma}} \quad \mathsf{id} \notin N_{M_{\sigma}}
	}
	& \hspace{10mm} &
	\infer[\mbox{E-ID2}]{E(\sigma, \mathsf{id}) \longrightarrow \langle \sigma,  (N_{\sigma_{M}} \mathsf{id}) ) \rangle}{%
		\mathsf{id} \in N_{M_{\sigma}}
	}
	\end{array}
	\]
	
	\[
	\footnotesize
	\infer[\mbox{E}_{RV}]{E_{R}(\sigma, \mathsf{exp}) \longrightarrow \langle \widehat{\sigma},v\rangle }{%
		\mathsf{E(\sigma, exp) \longrightarrow (\widehat{\sigma}, addr)}
		& \mathsf{[addr]_{(ST\ exp)_{\widehat{\sigma}}}^{Size (Type_{\widehat{\sigma}}\ exp)} =} v 
	}
	\]
	
	\[\footnotesize
	\infer[\textsc{E-STRUCT}]
	{E(\sigma, \mathsf{exp.k}) \longrightarrow \langle \widehat{\sigma}, \mathsf{addr_s} + \ceil*{\mathsf{SizeR\ 0\ \langle T_0 \cdots (T_k - 1)\rangle}}^{(l, T_i) }  \rangle}	
	{
		\begin{tabular}{l}
			$E(\sigma, \mathsf{exp}) \longrightarrow \langle \widehat{\sigma}, \mathsf{addr}  \rangle$ \quad 
			$\mathsf{[addr]_{(ST\ exp)_{\widehat{\sigma}}}^{Size (Type_{\widehat{\sigma}}\ exp)} =} addr_s$
			\\
			$\mathsf{k \leq i \quad \mathsf{Type}_\sigma\ exp = \langle T_0 \cdots T_i  \rangle}$ 
		\end{tabular}
	}
    \]
    
    \[\footnotesize
    \infer[\textsc{E-STRUCT-ref}]
    {E(\sigma, \mathsf{exp.k}) \longrightarrow \langle \widehat{\sigma}, \mathsf{addr} + \ceil*{\mathsf{SizeR\ 0\ \langle T_0 \cdots (T_k - 1)\rangle}}^{(l, T_i) }  \rangle}	
    {
    	\begin{tabular}{l}
    	$E(\sigma, \mathsf{Exp}) \longrightarrow \langle \widehat{\sigma}, \mathsf{addr}  \rangle$ \\
    	$\mathsf{k \leq i \quad \mathsf{Type}_\sigma\ exp = ref\ \langle T_0 \cdots T_i  \rangle}$ 
    	\end{tabular}
    }
    \]
    
	\[
	\footnotesize
	\infer[\mbox{E-ARRAY}]{E(\sigma, \mathsf{Exp_{b}[Exp_{i}]}) \longrightarrow \langle \widehat{\sigma}', v \rangle}{
		\begin{array}{l}
		\mathsf{Type_\sigma\ Exp_b = T[n]} \\
		\mathsf{i\leq n}
		\end{array}
		& \begin{array}{l}
		E_{R}(\sigma, \mathsf{Exp_{i}}) \longrightarrow \langle \widehat{\sigma}, i \rangle \\
		E(\widehat{\sigma}, \mathsf{Exp}) \longrightarrow \langle \widehat{\sigma}', addr_b \rangle \\
		v = \mathsf{addr_b + (i*Size\ T)}
		\end{array}
	}
	\]
	
	\[
	\footnotesize
	\infer[\mbox{E-ARRAY-REF}]{E(\sigma, \mathsf{Exp_{b}[Exp_{i}]}) \longrightarrow \langle \widehat{\sigma}', v \rangle}{
		\begin{array}{l}
		\mathsf{Type_\sigma\ Exp_b = ref\ T[n]} \\
		\mathsf{i\leq n} 		
		\end{array}
		& \begin{array}{l}
		E_{R}(\sigma, \mathsf{Exp_{i}}) \longrightarrow \langle \widehat{\sigma}, \mathsf{i} \rangle \\
		E_{R}(\widehat{\sigma}, \mathsf{Exp}) \longrightarrow \langle \widehat{\sigma}', \mathsf{addr} \rangle \\
		v = \mathsf{addr_a + (i*Size\ T)}
		\end{array}
	}
	\]

	\[
	\footnotesize
	\infer[\mbox{E-ARRAY-LEN}]{E(\sigma, \mathsf{exp.length}) \longrightarrow \langle \widehat{\sigma}, \mathsf{len}\rangle}{%
		\mathsf{Type_\sigma exp =  Type [\,]}
		& E_R(\sigma, \mathsf{exp}) \longrightarrow \langle \widehat{\sigma}, \mathsf{len} \rangle
	}
	\]
	
	\[
	\footnotesize
	\infer[\mbox{E-ARRAY-LEN-ref}]{E(\sigma, \mathsf{exp.length}) \longrightarrow \langle \widehat{\sigma}, \mathsf{len}\rangle}{%
		\mathsf{Type_\sigma exp = ref\ Type [\,]}
		& E_R(\sigma, \mathsf{exp}) \longrightarrow \langle \widehat{\sigma}, \mathsf{addr} \rangle 
        & \mathsf{[{addr}]_{(ST\ exp)_{\widehat{\sigma}}}^{Size (Type_{\widehat{\sigma}}\ Type[])} =len} 
	}
	\]
	
	\[
	\footnotesize
	\infer[\mbox{E-D-ARRAY}]{E(\sigma, \mathsf{Exp_b[Exp_i]}) \longrightarrow \langle \widehat{\sigma}', a \rangle}{
		\begin{array}{l}		
		\mathsf{Type_\sigma\ Exp_b = Type []} \\
		\mathsf{[{addr_b}]_{(ST\ Exp_b)_{\widehat{\sigma}'}}^{Size (Type_{\widehat{\sigma}'}\ Exp_b)} =v} \\
		i\leq (v-1)
		\end{array}
		& \begin{array}{l}
		E(\widehat{\sigma}, \mathsf{Exp_b}) \longrightarrow \langle \widehat{\sigma}', \mathsf{{addr_b}} \rangle \\
		E_R(\sigma, \mathsf{Exp_i}) \longrightarrow \langle \widehat{\sigma}, i \rangle \\
		a = \mathtt{addr32}(\mathtt{HASH}(\mathtt{bytes32}(\mathsf{addr_b})) + i)
		\end{array}
	}
	\]
	
		\[
	\footnotesize
	\infer[\mbox{E-D-ARRAY-ref}]{E(\sigma, \mathsf{Exp_b[Exp_i]}) \longrightarrow \langle \widehat{\sigma}', a \rangle}{
		\begin{array}{l}		
		\mathsf{Type_\sigma\ Exp_b =ref\ Type []} \\
		\mathsf{[{addr_b}]_{(ST\ Exp_b)_{\widehat{\sigma}'}}^{Size (Type_{\widehat{\sigma}'}\ Type[])} = v} \\
		\mathsf{i}\leq (\mathsf{v}-1)
		\end{array}
		& \begin{array}{l}
		E_R(\widehat{\sigma}, \mathsf{Exp_b}) \longrightarrow \langle \widehat{\sigma}', \mathsf{{addr_b}} \rangle \\
		E_R(\sigma, \mathsf{Exp_i}) \longrightarrow \langle \widehat{\sigma}, \mathsf{i} \rangle \\
		a = \mathtt{addr32}(\mathtt{HASH}(\mathtt{bytes32}(\mathsf{addr_b})) + \mathsf{i})
		\end{array}
	}
	\]
	
    \[
	\footnotesize
	\infer[\mbox{E-MAPPING}]{E(\sigma, \mathsf{Exp_b[Exp_i]}) \longrightarrow \langle  \widehat{\sigma}', \mathsf{v} \rangle}{
		\begin {array}{l}
		\mathsf{v} = \mathtt{addr32}(\mathtt{HASH}(\mathtt{bytes32}( \mathsf{{addr_b}}) \cdot \mathtt{bytes32}(\mathsf{i}))) \\
		\begin{array}{ll}		
		\mathsf{Type_\sigma\ Exp_b = Mapping\ K\ T} \quad & 
		E( \widehat{\sigma}, \mathsf{Exp_b}) \longrightarrow \langle  \widehat{\sigma}', \mathsf{{addr_b}} \rangle \\
		\mathsf{Type_\sigma\ Exp_i = K} &
		E_R(\sigma, \mathsf{Exp_i}) \longrightarrow \langle  \widehat{\sigma}, \mathsf{i} \rangle
		\end{array}		
		\end{array}
	}
	\]
	
	\[
	\footnotesize
	\infer[\mbox{E-MAPPING-REF}]{E(\sigma, \mathsf{Exp_b[Exp_i]}) \longrightarrow \langle  \widehat{\sigma}', \mathsf{v} \rangle}{
		\begin {array}{l}
		\begin{array}{ll}		
		\mathsf{Type_\sigma\ Exp_b = ref\ Mapping\ K\ T} \quad & 
		E_R( \widehat{\sigma}, \mathsf{Exp_b}) \longrightarrow \langle  \widehat{\sigma}', \mathsf{{addr_b}} \rangle \\
		\mathsf{Type_\sigma\ Exp_i = K} &
		E_R(\sigma, \mathsf{Exp_i}) \longrightarrow \langle  \widehat{\sigma}, \mathsf{i} \rangle
		\end{array} \\				
		\mathsf{v} = \mathtt{addr32}(\mathtt{HASH}(\mathtt{bytes32}( \mathsf{{addr_b}}) \cdot \mathtt{bytes32}(\mathsf{i}))) 		
		\end{array}
	}
	\]

\[
\footnotesize
\infer[\mbox{I-FUN}]{\langle \sigma, \mathsf{id(Exp_1, \ldots, Exp_n) }  \rangle \longrightarrow \langle \widehat{\sigma}',. \rangle}{%
	\begin{array}{l}	
	\mathsf{\langle M_{\sigma'}.Push(N',\tau'), T_1 \, M \,\,p_1 = Exp_1; \ldots; T_n \, M \,\, p_n = Exp_n;\, Block}\rangle \rightarrow \langle \widehat{\sigma},. \rangle   \\
	\Gamma_p \mathsf{ id = Block } \quad \Gamma_t\mathsf{ id = ((T_1, \ldots, T_n)},\_) \quad M_{\widehat{\sigma}'}.\mathsf{Pop()}
	\end{array}
}
\]

\[
\footnotesize
\infer[\mbox{E-FUN}]{E(\sigma, \mathsf{id(Exp_1, \ldots, Exp_n) }) \longrightarrow \langle \widehat{\sigma}',v \rangle}{%
	\begin{array}{l}	
	\mathsf{\langle M_{\sigma'}.Push(N',\tau'), T_1 \,\, M\,\,p_1 = Exp_1; \ldots; T_n \,\, M\,\, p_n = Exp_n;\, T_r \,\, M\,\, p_r;\, Block}\rangle \rightarrow \langle \widehat{\sigma},. \rangle   \\	
	\mathsf{v = N_{\widehat{\sigma}} p_r \quad 
		\Gamma_p \mathsf{id = Block } \quad \Gamma_t\mathsf{ id = ((T_1, \ldots, T_n),(Type_r))} \quad M_{\widehat{\sigma}'}.\mathsf{Pop()}}
	\end{array}
}
\]

\[
\footnotesize
\infer[\mbox{RETURN}]{\langle \sigma, \mathtt{return} \,\,\,\mathsf{Exp} \,; \rangle \longrightarrow \langle \widehat{\sigma'}, . \rangle}{%
	\begin{array}{l}		
	E_R(\sigma, \mathsf{Exp}) \longrightarrow \langle \widehat{\sigma}, v \rangle 
	\end{array}
	& \begin{array}{l}		
	\mathsf{[{N}_{\widehat{\sigma}} p_r]_{(ST\ exp)_{\widehat{\sigma}'}}^{Size\ ({\tau}_{\widehat{\sigma}} p_r)} =} v
	\end{array}
}
\]

\[
\footnotesize
\infer[\mbox{E-FUN}_1]{\langle \sigma, \mathsf{id_1.id_2(Exp_1, \ldots, Exp_n)\mathtt{.value}\mathsf{(Exp_{eth})}.\mathtt{gas}\mathsf{(Exp_{gas})}}  \rangle \longrightarrow \langle \ddot{\sigma}', \mathsf{id_2(Exp_1, \ldots, Exp_n)} \rangle}{%
	\begin{array}{l}
	\mathsf{id_1: Contract} \mbox{ in address } \alpha \\
	\mathsf{id_2: } \,\,\mathtt{function} \,\, \mathsf{(Type_1, \ldots, Type_n )}
	\end{array}
	& \begin{array}{l}
	\Delta(\alpha) = \ddot{\sigma} \\
	\ddot{\Omega}' = \ddot{\Omega}.\mathtt{push}(\sigma)
	\end{array}
	& \begin{array}{l}
	E(\sigma, \mathsf{Exp_{eth}}) \longrightarrow \langle \sigma, m \rangle \\
	E(\sigma, \mathsf{Exp_{gas}}) \longrightarrow \langle \sigma, n \rangle \\
	\end{array}
}
\]

\[
\footnotesize
\infer[\mbox{E-FUN}_2]{\langle \sigma, \mathsf{id}\mathtt{.call.value}\mathsf{(Exp_{eth})\mathtt{.gas}\mathsf{(Exp_{gas})}}  \rangle \longrightarrow \langle \ddot{\sigma}', \mathtt{fallBack()} \rangle}{%
	\begin{array}{l}
	\mathsf{id: Contract} \mbox{ in address } \alpha
	\end{array}
	& \begin{array}{l}
	E(\sigma, \mathsf{Exp_{eth}}) \longrightarrow \langle \sigma, m \rangle \\
	E(\sigma, \mathsf{Exp_{gas}}) \longrightarrow \langle \sigma, n \rangle
	\end{array}
	& \begin{array}{l}
	\Delta(\alpha) = \ddot{\sigma}\\
	\ddot{\Omega}' = \ddot{\Omega}.\mathtt{push}(\sigma)
	\end{array}
}
\]

\end{document}